\documentclass[%
reprint,
 amsmath,amssymb,
aps,
]{revtex4-2}
\newcommand{\tx}{{\tilde{x}}}
\newcommand{\ty}{{\tilde{y}}}
\usepackage{graphicx}% Include figure files
\usepackage{dcolumn}% Align table columns on decimal point
\usepackage{bm}% bold math
\usepackage{hyperref}% add hypertext capabilities
\begin{document}

%\preprint{APS/123-QED}

\title{Surface second-harmonic from metallic nanoparticle configurations - a transformation optics approach}% Force line breaks with \\

\author{K. Nireekshan Reddy$^{1}$}
\email{kothakap@post.bgu.ac.il}
\author{Parry Y. Chen$^1$}%
\author{Antonio I. Fern\'andez-Dom\'inguez$^2$}
\author{Yonatan Sivan$^1$} 
\affiliation{%
$^1$Unit of Electro-Optic Engineering, Ben-Gurion University, Be'er-Sheva, 8410501, Israel \\
$^2$
Departamento de F\'isica Te\'orica de la Materia Condensada and Condensed Matter Physics Center (IFIMAC), Universidad Aut\'onoma de Madrid E-28049 Madrid, Spain
}%
%\affiliation{
%}%
%\affiliation{
%\affiliation{%


\date{\today}% It is always \today, today,

\begin{abstract}
We study surface second-harmonic generation (SHG) from a singular plasmonic structure consisting of touching metallic wires. We use the technique of transformation optics and relate the structure to a rather simpler geometry, a slab waveguide. This allows us to obtain an analytical solution to the problem, revealing rich physical insights. We identify various conditions that govern the SHG efficiency. Importantly, our analysis demonstrates that apart from the mode-matching condition, phase-matching condition is relevant even for this sub-wavelength structure. Furthermore, we identify a geometric factor which was not identified before. We support our analysis with numerical simulations.

\end{abstract}

%Use showkeys class option if keyword
\maketitle

\section{Introduction}\label{sec_intro}
Transformation optics (TO) is a theoretical tool recently developed that allows an unprecedented control over light propagation and confinement~\cite{Ward_Pendry_1996,TO_review_Alex_2012, TO_JOP_review}. TO relies on the form-invariance of Maxwell's equations under coordinate transformations to provide the connection between a given electromagnetic effect, coded into a geometric transformation, and the material parameters ($\varepsilon$,~$\mu$) required for its realization~\cite{Pendry_OE_2006,TO_JOP_review}. 

Recently, it was shown that most of the complexity associated with TO schemes can be avoided by using a special class of transformations called conformal transformations (CT). These are two-dimensional (2D) mappings which have a very convenient characteristic - they preserve the material and spectral characteristics of the original system~\cite{Alex_kissing_cyls_NL,TO_review_Alex_2012}. This enables the description of the optical response of a wide variety of geometries by cascading conformal transformations~\cite{Alex_kissing_cyls_NL,Yu_singularities,TO_review_Alex_2012,Antonio_kissing_spheres_PRL,TO_vdw_science}. The analytic predictions of TO have been found to be in excellent agreement with numerical results (for subwavelength particles), and verified experimentally by several groups~\cite{Lei_sphere_plane,Hanham_AdvMat,Ciraci:2012_Science,Baumberg-Aizpurua}. In addition, CT allowed addressing various other physical effects such as the moulding of surface plasmon polariton propagation~\cite{Huidobro_NanoLett}, the emergence of non-local effects in metallic nanostructures~\cite{Antonio_kissing_cyls_nonlocal_PRL,Antonio_kissing_cyls_nonlocal_PRB}, van der Waals interactions at the nano-scale~\cite{Zhao_PRL}, graphene~\cite{Huidobro_ACSNano} and singular~\cite{Paloma_science} metasurfaces, and plasmon-exciton interactions~\cite{Cuartero_ACSPhot} and strong coupling~\cite{RiuQi_PRL}.  

\begin{figure}[b!]
\begin{center}
\includegraphics[width=\columnwidth]{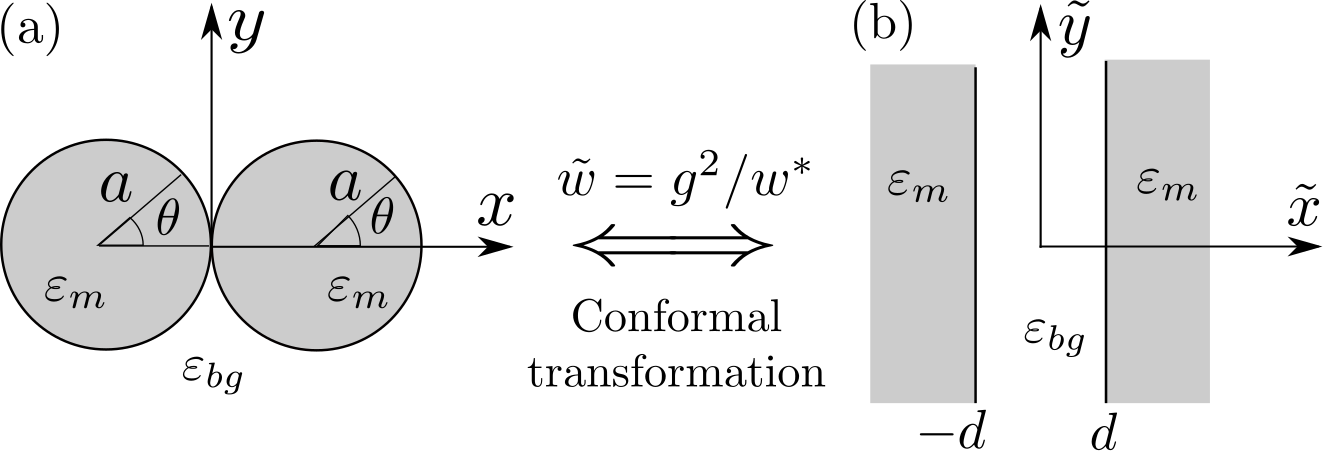}
\caption{Schematics of the (a) identical touching wire system and the (b) slab geometry related through conformal transformation. % add incident plane wave? 
}
\label{f1}
\end{center}
\end{figure}

The most fundamental contribution of CT is, perhaps, to establish the equivalence between the (linear) scattering from isolated particles (supporting localized plasmon resonances) and light propagation along straight waveguides (supporting surface plasmon resonances)~\cite{Alex_kissing_cyls_NL,Antonio_kissing_spheres_PRL,TO_vdw_science,DouglasWerner_OE}, at least within the quasi-static approximation. This enabled calculation of the field distribution in complex structures and especially the challenging singular geometries by first transforming to simple, regular solvable geometries, and then transforming back (see Figure~\ref{f1}). Such geometries give rise to extremely high local-field enhancement near the singularities. The CT also showed that the spectrum of the singular geometry is ``inherited'' from the broadband plasmonic waveguide structure, thus, resolving the long-standing arguments about the origin of the broadband spectral response from rough (disordered) and/or multiscale structures~\cite{Alex_kissing_cyls_NL,mcphedran_1987_TW}. This class of structures thus allow for breaking the conventional resonance-based limitations of bandwidth in plasmonic devices. 
To date, {\em all} the studies employing TO were limited to media that have a {\em linear} response to the incoming electromagnetic field, namely, for low intensities~\footnote{The only exception is a paper that focused on the general formulation and provided a single example of third-order nonlinearity~\cite{NL_TO}. }. However, extending TO to nonlinear wave interactions is appealing, since both the high local-field enhancement as well as the unusually wide bandwidth make the singular structures potentially useful for such interactions.

Here, we employ TO for a singular nanoparticle configuration that includes {\em nonlinear} media (specifically, for Second Harmonic Generation, SHG from metal nanoparticles). We demonstrate the strength of our approach via an analytic solution of the near-field distribution and conversion efficiency, which is found to be in excellent agreement with exact electrodynamic numerical simulations. Most importantly, we provide deep insights of the physics underlying frequency conversion processes.

In particular, we show that the theoretical description of nonlinear wave mixing from nanoparticles is more complex than one would have expected compared to the metal waveguides. Indeed it is well known that in the latter case, efficient SHG has three requirements - a strong source, phase matching (PM) and mode matching (MM)~\cite{Khurgin_SHG_MDM,Kivshar_SHG_MDM}. Yet, for subwavelength structures, e.g., single nanoparticles or nanoparticle clusters, it is customary to dismiss the need to achieve phase matching~\cite{zayat_rev}, justified by the claim that the phase accumulation across the structure is small, so that the source and generated wave are nearly unaffected by any phase mismatch. This implies that there is a fundamental difference between extended structures (such as waveguides) and particle geometries. However, the intrinsic analogy between the infinitely extended flat geometry and particle(s) geometry (see Figure~\ref{f1})~\cite{TO_review_Alex_2012} implies that this is {\em not} justified in all cases%{\bf(Antonio: I think we must not generalize this result beyond our peculiar geometry)}
. Indeed, the CT map shows that the modes propagating an infinite distance in the flat geometry are the same modes that repeatedly circulate around the touching wires (TWs). Thus, since it is trivial that PM is essential in the flat geometry, then, it is necessarily as important in the particle geometry. Indeed, particle-induced wavelength compression makes the accumulated phase substantial despite the short propagation distance along the circumference and regardless of any detuning. {\em In that sense, a critical observation is that for the purpose of frequency conversion, the particle geometry is very different from a bulk material or homogenized composite of the same dimensions} for which the phase mismatch is not important.

Below, we show explicitly that PM does manifest itself in small NPs, and connect it to the localized plasmon resonance%and the so-called selection rules~\cite{Selection_rules}
. Then, we show, somewhat unexpectedly, that these so-called three requirements are not sufficient, and that in practice there is an additional consideration. We discuss how this additional term modifies the  SH response.

The article is organized as follows: Section~\ref{sec:formult+method} describes the identical TW system, SH surface sources and their symmetry aspects and formulates the SHG problem in the TW frame. To obtain the SHG solution, we follow the route of conformal transformation to transform TW geometry to a rather simpler geometry, i.e., slab geometry and solve for SHG in Section~\ref{sec:trsfm_to_slab} . We then transform the obtained solution back to the TW geometry in Section~\ref{sec:back_transfm} and interpret it. Finally, we conclude with a discussion and outlook.

\section{Touching wire system and surface nonlinear sources}\label{sec:2}

\subsection{Formulation and methodology}\label{sec:formult+method}

Consider two identical metallic TWs of radii $a$ with permittivity $\varepsilon_m$, touching each other at the coordinate origin of the $x-y$ plane; they are embedded in a homogeneous dielectric background of permitivitty $\varepsilon_{bg}$ (see Figure~\ref{f1}(a)). The TW structure is assumed to be illuminated by a TM polarized plane wave at a frequency $\omega$ by a spatially uniform $x$-polarized electric field with amplitude $E^{\omega}_{0x}$. The linear electric field response of the structure at the fundamental frequency (FF), $\mathbf{E}^\omega$, can be evaluated analytically using the technique of conformal transformation as in~\cite{Alex_kissing_cyls_NL,alex_kissing_NJP}. Firstly, the TW geometry and the plane wave source are transformed to a metal-dielectric-metal slab structure and a line dipole source, respectively. Then, the dipole radiation is coupled to plasmonic modes of the slab system in the momentum space. The dominant contribution to the linear response arises from the surface plasmon pole and the contribution from the lossy surface (radiative) waves was neglected. Having obtained the closed form solution in the slab frame, the solution was then transformed back to the TW frame. The analytic solution revealed several interesting physical phenomena occurring close to the singular (touching) point. Most notable of them are the unusually wide spectral response, strong spatial field confinement and large field enhancement close to the touching point. The latter effect can give rise to efficient nonlinear optical phenomena.

In this article, we consider a second-order process, specifically, second-harmonic generation (SHG). It is well known that the second-order nonlinear processes are symmetry forbidden in centro-symmetric materials. However, due to the broken symmetry at an interface, the metal-dielectric interface has a non-zero second-order surface tensor $\bar{\bar{\chi}}^{(2)}_S$ \cite{rudnik,Heinz_thesis}. Apart from purely surface effects, non-locality can give rise to nonlocal bulk nonlinearity which can be mapped to a surface current source, so that the surface nonlinearity provides a simple, general model for second-order nonlinear phenomena in metals~\cite{sipe_srf_map,miano,ciraci_prb_SHG,ciraci2}. We assume that the surface SH polarization $\mathbf{P}^{2\omega}_S$ at the metal-dielectric interface is given by
\begin{equation}
P^{2\omega}_{S,\perp} = \varepsilon_0\chi^{(2)}_{S,\perp\perp\perp} E_\perp^{\omega} E_\perp^{\omega}~\delta(x^2+y^2\mp2ax), \label{SHpol}
\end{equation}
where $E_\perp^{\omega}$ and $P^{2\omega}_{S,\perp}$ correspond to the normal component (to the metal-dielectric interface) of linear electric field and surface SH polarization, respectively. The metal-dielectric interface of the right (left) wire corresponds to $-$ ($+$) sign in the argument of the Dirac-delta function $\delta$. 

In writing $P^{2\omega}_{S,\perp}$, we assume that the dominant contribution to the surface nonlinear polarization arises from the $\chi^{(2)}_{S,\perp\perp\perp}$ element, see justification in Ref.~\cite{jnl}. To obtain $P^{2\omega}_{S,\perp}$, we use the analytical expressions of the FF electric field $E^\omega_\perp$ derived in Ref.~\cite{alex_kissing_NJP}. We note that $E^{\omega}_{\perp}$ is discontinuous across the interface, thus, leading to an ambiguity in choosing $E^{\omega}_{\perp}$ either on the metal or dielectric side of the interface. Following Sipe \textit{et al.}~\cite{sipe80}, we choose $E^\omega_\perp$ on the metal side of the interface. 

In the solution of SH fields, Eq.~\eqref{SHpol} appears as a source term in Maxwell's equations, and can be solved in this form~\cite{knr_josab,chen2017generalizing}. But in surface SHG, the sources coincide with the boundary, so alternatively $P^{2\omega}_{S,\perp}$ can be incorporated into the boundary conditions at the metal-dielectric interface~\cite{Heinz_thesis}. We shall adopt this convenient approach. 

First, note that the SH polarization $P^{2\omega}_{S,\perp}$ is normal to the interface, corresponding to an electric dipole layer pointing normal to the interface. In such a case, the tangential component of the SH field $\mathbf{E}_\parallel^{2\omega}$ is discontinuous. The discontinuity across the interface, denoted as $\Delta \mathbf{E}_\parallel^{2\omega}$, is given by the generalized boundary condition as $\Delta \mathbf{E}_\parallel^{2\omega} = - \frac{1}{\varepsilon_{bg}^{2\omega}}\nabla_\parallel P^{2\omega}_{S,\perp}$, where $\nabla_{\parallel}$ is the tangential derivative along the interface ~\cite{Heinz_thesis,knr_josab}. This generalized boundary condition can be conveniently reformulated using the divergence-free magnetic surface current density $\mathbf{J}_{mS}^{2\omega}$~\cite{martti_bem,knr_josab}. In such a case, the generalized boundary condition takes the form \begin{eqnarray}
\Delta \mathbf{E}_{\parallel}^{2\omega}=-\mathbf{n}\times\mathbf{J}_{mS}^{2\omega},~\textrm{with}~ \mathbf{J}_{mS}^{2\omega}= \frac{1}{\varepsilon_{bg}^{2\omega}} \mathbf{n}\times \left(\nabla_{\parallel}P^{2\omega}_{S,\perp}\right), \label{genbc} 
\end{eqnarray}
where $\mathbf{n}$ is the unit normal to the interface. The magnetic surface current density generated by the SH polarization $P^{2\omega}_{S,\perp}$ considered in Eq.~\eqref{SHpol} for the TW geometry is given by
\begin{equation}
J_{z,r/l}(x,y)=\frac{\chi^{(2)}_{S,\perp\perp\perp}}{\varepsilon_{bg}^{2\omega}}~\partial_{\parallel} (E_\perp^{\omega}E_\perp^{\omega}) ~\delta(x^2+y^2\mp2ax),\label{jmz}
\end{equation}
where $J_{z}$ is the out-of-plane (i.e., $z$) component and furthermore, the only component of the magnetic current $\mathbf{J}_{mS}^{2\omega}$ and we have dropped the superscript $2\omega$ and the subscript $mS$ of $J_{mS,z}^{2\omega}$ for the sake of brevity. The subscript $r$($l$) of $J_{z,r/l}$ denotes the right(left) wire and $\partial_{\parallel}$ denotes the tangential derivative along the interface.

\begin{figure}[h!]
\begin{center}
\includegraphics*[height=4.5cm, width=\columnwidth]{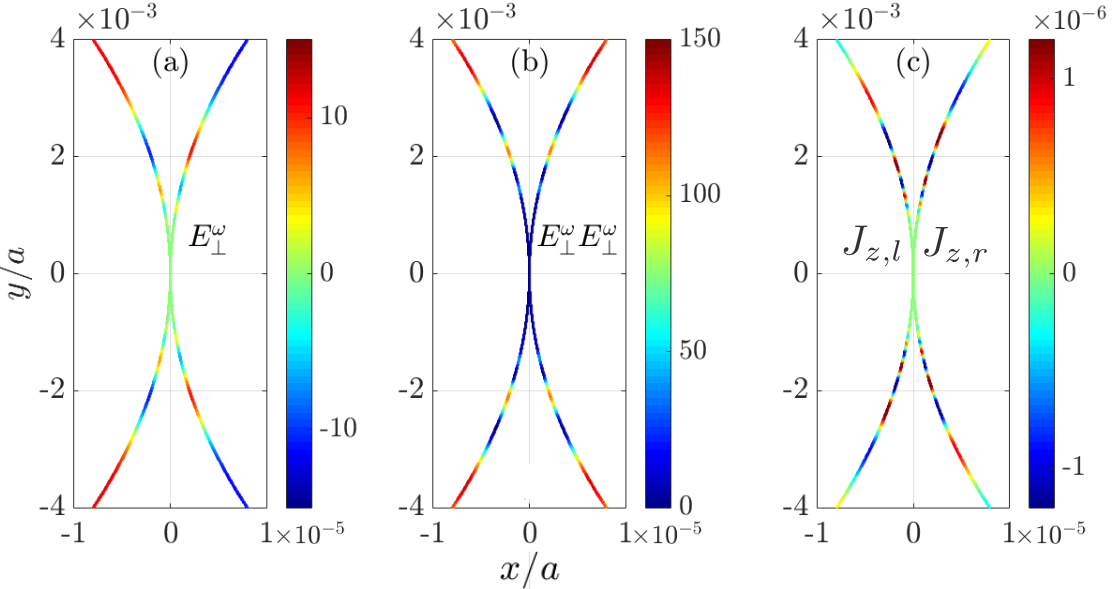}
\caption{(Color online) Various fields plotted on the circumference of the touching wires close to the touching point. (a) $E_\perp^{\omega}$ (below the surface plasma resonance frequency) is anti-symmetric in $x$ and symmetric in $y$. (b) $E_\perp^{\omega}E_\perp^{\omega}$ is symmetric in $x$ and $y$. (c) $J_{z,r/l}$~\eqref{jmz} is anti-symmetric in $x$ and $y$, thus, $J_{z,r} = - J_{z,l}$. All distributions plotted for $\omega = $. }
\label{fig_jmz} 
\end{center}
\end{figure}

Since our TW structure and the incidence is symmetric, we now investigate the symmetry relations of the $J_{z,r/l}$ on the left and right wires. Figure~\ref{fig_jmz} shows various fields (on the metal side) on the circumference of the TW close to the touching point. As shown in Figure~\ref{fig_jmz}(a), the linear response $E_\perp^\omega$ has a definite symmetry which is anti-symmetric in $x$ and symmetric in $y$ below the surface plasma resonance frequency. Since $E_\perp^\omega E_\perp^\omega$ is quadratic in $E_\perp^\omega$, then necessarily $E_\perp^\omega E_\perp^\omega$ is symmetric in $x$ and $y$ (see Figure~\ref{fig_jmz}(b)), as expected. The tangential derivative $\partial_{\parallel}$ of $E_\perp^\omega E_\perp^\omega$ along the circumference of the left and right wires yields $J_{z,r/l}$~\eqref{jmz} which is an anti-symmetric source in $x$ and $y$. As a consequence, $J_{z,r}(x,y)= - J_{z,l}(-x,y)$ (see Figure~\ref{fig_jmz}(c)). Note that the SH source symmetry described above holds well for the choice of $\chi^{(2)}_{S,\perp\perp\perp}$ element at any frequency of operation and all material parameters.

Having understood the SH source and its symmetries, we now formulate the SH problem for the TW. Since the SH response is usually weak when compared to the FF response, we employ the undepleted pump approximation~\cite{Boyd-book}. Furthermore, we also assume the physical dimensions of the wire $a$ to be much smaller than both the FF and SH wavelengths such that one can safely employ the quasi-static approximation (namely, we take the limit $(\omega/c)a \ll1$). Since we consider in-plane $TM$ polarized illumination, it is convenient to formulate a scalar SHG problem in terms of the SH magnetic field $H_{z,TW}^{2\omega}$. We emphasize that for a quasi-static structure, the magnetic response is rather weak yet non-zero (see Appendix~\ref{apen:lin_mag} for further discussion). Upon obtaining the SH magnetic field, the corresponding SH electric fields can then be evaluated using Ampere's law. 

The SH magnetic field $H_{z,TW}^{2\omega}$ can be obtained by solving the homogeneous Helmholtz equation in each domain accompanied with two boundary conditions. However, in the quasi-static limit Helmholtz equation in each domain simplifies to the homogeneous Laplace equation given by
\begin{equation}
\nabla ^{2} H_{z,TW}^{2\omega}=0. \label{hlmhz_qs}
\end{equation}
Futher, the two boundary conditions are as follows. Continuity of $H_{z,TW}^{2\omega}$ across the interface is given by
\begin{equation}
\mathbf{n}\times(\mathbf{H}_{bg}-\mathbf{H}_{m})=H_{z,TW}^{2\omega}|_{bg}-H_{z,TW}^{2\omega}|_{m}=0,\label{bc_1}
\end{equation}
where the subscripts $bg$ and $m$ correspond to the dielectric and metal sides of the interface, respectively. The generalized boundary condition (see Eq.~\eqref{genbc}) in terms of $H_{z,TW}^{2\omega}$ and the magnetic surface density $J_{z,r/l}$ for the TW is given by
\begin{equation}
\left[\mathbf{n}\cdot \nabla\left(\frac{ H_{z,TW}^{2\omega}|_{bg}}{\varepsilon_{bg}^{2\omega}}-\frac{ H_{z,TW}^{2\omega}|_{m}}{\varepsilon_m^{2\omega}}\right)\right]=-2i\omega\varepsilon_0J_{z,r/l}.\label{bc_2}
\end{equation}
Invoking the definition of $\mathbf{n}$ for the right and left wires, Eq.~\eqref{bc_2} can be rewritten as
\begin{equation}
\left(\frac{x\pm a}{a}\partial_{x} +\frac{y}{a} \partial_{y} \right)\left[\frac{ H^{2\omega}_{z,TW}|_{bg}}{\varepsilon_{bg}^{2\omega}} -\frac{ H^{2\omega}_{z,TW}|_{m}}{\varepsilon_m^{2\omega}} \right]   
= -2i\omega\varepsilon_0 J_{z,r/l}. \label{full_gnbc} 
\end{equation}

We solve the SH problem analytically using conformal transformation by relating the TW to a much simpler geometry, a slab waveguide, as for the linear case~\cite{Alex_kissing_cyls_NL}.

\subsection{Transformation to the slab frame} \label{sec:trsfm_to_slab} 
Consider the conformal coordinate transformation 
\begin{equation}
\tilde{w}=\frac{g^2}{w^*}. \label{eq_CT}
\end{equation}
Here, we use the complex number notation $w = x + iy$ for the TW coordinates and $\tilde{w} = \tx + i \ty$ for the new coordinate system ($*$ denotes the complex conjugate). The parameter $g$ in Eq.~\eqref{eq_CT} is a scaling constant with dimensions of length. The coordinates of the old and new systems are related by
\begin{equation}
\tx= \frac{g^2x}{x^2+y^2},\quad \ty= \frac{g^2y}{x^2+y^2}. \label{eq_xx'} 
\end{equation}
The touching wire system (in $x-y$ plane) transforms to a metal-dielectric-metal slab waveguide (in $\tx-\ty$ plane) of half-width $d=g^2/2a$ (see Figure~\ref{f1}). The permitivitty of the materials remains unchanged under this 2D conformal transformation \cite{alex_kissing_NJP}. 

Having transformed the TW system to slab frame, we now formulate the SH problem in the slab geometry by transforming the governing equation \eqref{hlmhz_qs} and the two boundary conditions Eqs.~\eqref{bc_1} and \eqref{full_gnbc}.
Firstly, we note that the magnetic field is invariant under 2D conformal transformation (see Appendix~\ref{apen:Hz_transfrm}) as it is an out-of-plane component~\cite{Ward_Pendry_1996,Huidobro_JOP_2016}. Therefore, $H_{z,TW}^{2\omega}(x,y) = H_{z,sl}^{2\omega}(\tx(x,y),\ty(x,y))$). Now transforming Eqs.~\eqref{hlmhz_qs}-\eqref{bc_1} and expressing them in terms of the slab variables, we arrive at
\begin{equation}  \label{hlhmtz_sl}
[\partial^2_{\tx}+\partial^2_{\ty}] H_{z,sl}^{2\omega} = 0\,,  \end{equation} 
\begin{equation}
H_{z,sl}^{2\omega}|_{bg} = H_{z,sl}^{2\omega}|_{m}\,, \label{bc_1_sl} 
\end{equation} 
respectively. 

Before transforming the boundary condition in Eq.~\eqref{full_gnbc}, let us examine $J_{z,r/l}$ in the TW frame. We decompose $J_{z,r/l}$ in terms of amplitude and phase as 
\begin{subequations}
\begin{align} 
\label{jmz_decmpr}
J_{z,r}(x,y) &= \mathcal{R}(x,y)~\exp{ \left( \frac{4ia\alpha^\omega|y|}{x^2+y^2} \right)} ~\delta(x^2+y^2-2ax), \\
J_{z,l}(x,y) &= \mathcal{L}(x,y)~\exp{ \left(\frac{4ia\alpha^\omega|y|}{x^2+y^2} \right)} ~\delta(x^2+y^2+2ax), \label{jmz_decmpl}
\end{align} \label{jmz_decmp}
\end{subequations} 
\noindent where $\mathcal{R}$ and $\mathcal{L}$ correspond to the amplitude of the SH source on the right and left wires, respectively (see Appendix~\ref{apen:RandL}). We now describe in detail the origin of the functional dependence of the phase part of $J_{z,r/l}$ in Eqs.~\eqref{jmz_decmp}.

Recall that the linear response was analytically obtained by transforming the TW geometry to the slab~\cite{Alex_kissing_cyls_NL}. This slab structure in the quasi-static limit exhibits two surface plasmon modes. The symmetric mode extends from zero frequency to the surface plasmon resonance, while the anti-symmetric mode exists above the surface plasmon resonance. The linear solution involved only the symmetric mode whose dimensionless (i.e., the product of the propagation constant and slab half-width) propagation constant at FF is given by~\cite{Maier-book} \footnote{Lei \textit{et al.}\cite{alex_kissing_NJP} consider a slab of width $d$. Therefore, their dimensionless propagation constant differs by a factor 2 from ours. We adopt this change of notation in order to conform to standard PM notations, see below.}
\begin{equation}
\alpha^\omega = \frac{1}{2}\log\left(\frac{\varepsilon_m^\omega-\varepsilon_{bg}^\omega}{\varepsilon_m^\omega + \varepsilon_{bg}^\omega}\right)  \quad  \textrm{when ~~Re}(\varepsilon_m)<-\varepsilon_{bg}^\omega. \label{alp}
\end{equation}
Despite transforming the FF solution back to the TW frame, $\alpha^\omega$ still continues to act as the propagation constant for the TW system~\cite{hidden_symm_2014}. As a consequence, the phase of the linear field $E_{\perp}^{\omega}$ on the circumference of the identical TW is given by $\exp{ \left(2i a \alpha^\omega |y|/(x^2 + y^2)\right)}$  \cite[Eqs.~(32)-(37)]{alex_kissing_NJP}. Since the SH source is obtained upon squaring the linear fields (see Eq.~\eqref{jmz}), the phase of SH source is exactly twice the FF phase. Furthermore, as $J_{z,r}(x,y) = - J_{z,l}(-x,y)$, we have $\mathcal{R}(x,y) = - \mathcal{L}(-x,y)$.

The CT unfolds the considerably complicated SH source $J_{z,r/l}$ to a much simpler one in the slab frame. Specifically, transforming $J_{z,r/l}$~\eqref{jmz_decmp} to the slab frame yields (see Appendix~\ref{apen:gnbc_transfrm}) \begin{equation}
J_{z,r/l}~ \rightarrow ~\Delta_{r/l}(\tx,\ty)~ e^{2i\alpha^\omega|\ty|/d} ~\delta(\tx = \pm d),
\label{jmz_delrt}
\end{equation}
where $\Delta_r$ and $\Delta_l$ are the amplitudes of the magnetic surface currents in the slab geometry obtained by transforming $\mathcal{R}$ and $\mathcal{L}$, respectively. Thus, $\Delta_r(\tx = d,\ty) = -\Delta_l(\tx = - d,\ty)$ as $\mathcal{R}(x,y) = -\mathcal{L}(-x,y)$, the phase part of $J_{z,r/l}$ (see Eqs.~\eqref{jmz_decmp}) simplifies to $e^{2i\alpha^\omega|\ty|/d}$ with $2\alpha^\omega$ as dimensionless propagation constant of the SH source in the slab frame, and the SH surface source $J_{z,r}$ ($J_{z,l}$) on the circumference of the right (left) wire transforms to line source placed at the right (left) interface of the slab.

The generalized boundary condition~\eqref{full_gnbc} now transforms to the slab frame as (see Appendix~\ref{apen:gnbc_transfrm}) 
\begin{widetext}
\begin{subequations}
\begin{align}
\left(\frac{d^2 + \ty^2}{g^2}\right) \partial_{\tx} \left[\frac{H_{z,sl}^{2\omega}|_{bg}}{\varepsilon_{bg}^{2\omega}} - \frac{H_{z,sl}^{2\omega}|_{m}}{ \varepsilon_m^{2\omega}}\right] &=  2 i \omega \varepsilon_0~\Delta_r(\tx=d,\ty)~e^{2i\alpha^\omega|\ty|/d},\quad \textrm{at}~\tx=d,\label{hnztra} \\
\left(\frac{d^2 + \ty^2}{g^2}\right) \partial_{\tx} \left[ \frac{H_{z,sl}^{2\omega}|_{bg}}{\varepsilon_{bg}^{2\omega}} - \frac{H_{z,sl}^{2\omega}|_{m}}{ \varepsilon_m^{2\omega}}\right] &= - 2 i \omega \varepsilon_0 \Delta_{l}(\tx=-d,\ty)~e^{2i\alpha^\omega|\ty|/d}, \quad \textrm{at}~\tx = -d. \label{hnztrb} 
\end{align}
\label{hnztr}
\end{subequations}
\end{widetext}
We note that the spatial factor $(d^2 + \ty^2)$ on the left-hand-side of Eqs.~\eqref{hnztr} arises from transforming $\mathbf{n}\cdot\nabla$; this term thus encodes information of the original TW geometry (see Eqs.~\eqref{bc_2}-\eqref{full_gnbc}).

We can now calculate the SHG response in the slab geometry by solving Eq.~\eqref{hlhmtz_sl} with boundary conditions in Eq.~\eqref{bc_1_sl} and Eqs.~\eqref{hnztr} using an \emph{ansatz}. The mathematical form of our \emph{ansatz} for $H_{z,sl}^{2\omega}$ is dictated by the governing physical phenomena, namely,
\begin{enumerate}
\item The ansatz should satisfy the Laplace equation~\eqref{hlhmtz_sl} in all the domains.

\item Since we are interested in the near-field of the nanostructure which is expected to be governed by surface plasmon waves rather than photonic waveguide modes, the \emph{ansatz} should exhibit exponential decay along the transverse coordinate $\tx$ away from the interfaces at $\tx = \pm d$.

\item Since the slab structure and SH source have definite symmetry in the $\tx-\ty$ plane, the \emph{ansatz} should also have definite a symmetry. In fact, the anti-symmetric magnetic line sources at the interfaces ($\Delta_r = - \Delta_{l}$) dictates that the \emph{ansatz} should be anti-symmetric in $\tx$. 

\item The longitudinal dependence of the \emph{ansatz} should be the same as the SH source, i.e., $e^{2i \alpha^\omega|\ty|/d}$ as it is specified for all $\ty$.

\item Having set the propagation constant of the \emph{ansatz} along $\ty$ to be $\alpha^\omega$, the dispersion relation gives us the propagation constant along the transverse direction $\tx$ as $k_{\tx, bg/m} = 2\sqrt{(\alpha^\omega)^2 - \left( \frac{\omega}{c}\right)^2 \varepsilon^{2\omega}_{bg/m}}$. However, we operate in the quasi-static regime, i.e., in the limit $\omega/c \rightarrow 0$, therfore $k_{\tx, bg/m}$ simplifies to $2\alpha^\omega$, making the transverse propagation constant the same as that of the longitudinal one in all the regions.
\end{enumerate}

\bigskip

Following the above considerations and assuming that $H_{z,sl}^{2\omega}$ is variable separable, we arrive at the \emph{ansatz}
\begin{eqnarray}
H_{z,sl}^{2\omega}(\tx,\ty) = \begin{cases}
\mathcal{A}_{sl}~e^{2i\alpha^\omega |\ty|/d}\sinh\left(2\alpha^\omega \tx /d\right) ,~~ |\tx|\leq d, \\ 
\mathcal{B}_{sl}~e^{2i\alpha^\omega |\ty| /d} ~\textrm{sgn} [\tx]  e^{-2\alpha^\omega |\tx| /d} ,~~ |\tx|>d.
\end{cases} \label{ansatz_sl} 
\end{eqnarray}
The amplitudes $\mathcal{A}_{sl}$ and $\mathcal{B}_{sl}$ in Eq.~\eqref{ansatz_sl} are the slowly varying amplitudes~ \footnote{The \emph{ansatz}~\eqref{ansatz_sl} satisfies Eq.~\eqref{hlhmtz_sl} only when the amplitudes $\mathcal{A}_{sl}$ and $\mathcal{B}_{sl}$ are slowly varying such that their spatial derivatives can be neglected.} in
$\ty$ such that the \emph{ansatz} satisfies Eq.~\eqref{hlhmtz_sl} and Eqs.~\eqref{hnztr}. 

Demanding the continuity of $H_{z,sl}^{2\omega}$ across the interfaces (see Eq.~\eqref{bc_1_sl}) gives
\begin{equation}
\mathcal{B}_{sl}= \mathcal{A}_{sl}~e^{2\alpha^\omega}\sinh \left(2\alpha^\omega\right) \,.
\label{sl_bc1_impl}  
\end{equation}
Implementing the transformed generalized boundary condition at the right and left interfaces (see Eq.~\eqref{hnztr}) and using Eq.~\eqref{sl_bc1_impl} yields
\begin{gather}
\mathcal{A}_{sl} = \frac{-i \omega \varepsilon_0 \varepsilon_{bg}^{2\omega} g^2 d}{\alpha^\omega \mathcal{P}} ~\left(\frac{\Delta_r(\tx = d,\ty)}{d^2 + \ty^2}\right), \label{sl_bc2_impl} \\ 
\mathcal{P} = \cosh \left(2\alpha^\omega\right) + \frac{\varepsilon^{2\omega}_{bg}}{ \varepsilon^{2\omega}_m} \sinh \left(2\alpha^\omega\right)\,. \label{eqn_P}
\end{gather}
Let us now interpret the SH solution~(\ref{ansatz_sl})-(\ref{eqn_P}). First, we note that the spatial dependence of the solution~(\ref{ansatz_sl}) has a striking resemblance to an anti-symmetric mode of the slab~\cite{Maier-book,Kivshar_SHG_MDM,Fan_THz_MDM,Khurgin_SHG_MDM}. However, potentially unexpectedly (e.g., based on analysis such as coupled mode theory, see e.g.,~\cite{Kivshar_SHG_MDM,Khurgin_SHG_MDM,Pavel_NJP}), it does not necessarily correspond to the excitation of a single mode of the structure because the phase accumulation rate (i.e., the $\ty$ dependence) is determined by the source rather than by a propagation constant of a given (\emph{anti}-symmetric) mode; in the quasi-static limit, these dimensionless propagation constants $\alpha^{2\omega}$ are given by the roots of the dispersion relations~\cite{Maier-book}\footnote{For the electrodynamic case, a root search is required~\cite{davidenko,chen2017robust}.}
\begin{equation}
\cosh \alpha^{2\omega} + \frac{\varepsilon^{2\omega}_{bg} }{\varepsilon^{2\omega}_m} \sinh \alpha^{2\omega} = 0, \label{disp_antisymm}
\end{equation}
which has solutions only for 
\begin{equation}
-\varepsilon^{2\omega}_{bg} < \textrm{Re}(\varepsilon^{2\omega}_m) < \varepsilon^{2\omega}_{bg}, \label{antisymm}   
\end{equation}
which is above the surface plasmon resonance frequency.

A remarkable aspect of conformal transformation is that the source exhibits harmonic phase variation along $\ty$ (see Eq. \eqref{jmz_delrt}), which is preserved in the solution via momentum conservation. This also enables elegant phase matching interpretations. Firstly, one can appreciate that the absence of propagation constants of the SH modes from the solution~(\ref{ansatz_sl}) is in fact expected, since the configuration we study is analogous to the scattering problem from a Fabry-P\'erot etalon. Indeed, the spatial profile of the electric fields is known in each domain, and the corresponding unknown amplitudes are calculated by ensuring the satisfaction of the boundary conditions. In that procedure, the scattered fields have the same longitudinal momentum as that of the source (incident field) and the excitation is resonant only when the source momentum matches that of a mode of the etalon. 

We now employ the powerful phase matching interpretation of the result. A SH mode is resonantly excited in the slab only if the longitudinal momentum component of the SH source matches the momentum of the (anti-symmetric) mode. This can be inferred by comparing Eq.~\eqref{disp_antisymm} to $\mathcal{P}$ in Eq.~\eqref{eqn_P} where we see that PM is obtained when $2\alpha^{\omega} = \alpha^{2\omega}$. Thus, the above reveals the analogy between a resonant excitation and PM for SHG. Further, we note that PM results in the enhancement of the SH fields as $H_{z,sl}^{2\omega}$ diverges for $\mathcal{P} \rightarrow 0$. Unfortunately, a comparison of the expressions of $\alpha^\omega$ (Eq.~\eqref{alp}) and $\alpha^{2\omega}$ (Eq.~\eqref{disp_antisymm}) shows that due to material dispersion, the zeros of $\mathcal{P}$ cannot be reached without special means (see e.g.,~\cite{Kivshar_SHG_MDM,Khurgin_SHG_MDM}); in practice the values taken by $\mathcal{P}$ are even greater in the presence of absorption, as these push the poles to the complex plane of the parameter domain.

An additional condition for efficient SHG usually found in the literature is the so-called mode matching (MM) condition. In contrast to PM discussed above, which essentially involves the spatial overlap of the source and mode in the \emph{longitudinal} coordinate, MM refers to the spatial overlap of the mode and source in the direction \emph{perpendicular} to the propagation; maximal overlap is desired to achieve efficient conversion. For symmetric structures, the overlap integral has a ``binary'' interpretation - a symmetric (anti-symmetric) source can only excite the symmetric (anti-symmetric) modes. Furthermore, for the line sources arising from the surface nonlinearity studied here, the overlap simplifies drastically to the product of the amplitudes on each of the interfaces. As explained, in our case, the anti-symmetric combination of the SH line sources at the interfaces ($\Delta_r$ and $\Delta_l$) dictates that the SH \emph{ansatz}~\eqref{ansatz_sl} has to be anti-symmetric along the transverse coordinate $\tx$. As a consequence, the condition $-\varepsilon_{bg}^{2\omega} < \textrm{Re}(\varepsilon_m^{2\omega}) < \varepsilon_{bg}^{2\omega}$ in Eq.~\eqref{antisymm} reveals that PM can be achieved only for frequencies above the surface plasmon resonance frequency. Since the modes occurring above the surface plasmon resonance frequency are very lossy, the SH efficiency is expected to be low.

All the above clarifies the role of phase-matching and mode-matching conditions in the slab geometry for SHG with surface sources. Apart from the above, we also note an additional feature in our solution when compared to the usual slab solution, namely, the term $(d^2 + \ty^2)$ in the denominator of the SH solution~\eqref{sl_bc2_impl}. As mentioned earlier, it originates from the transformation of the boundary condition~\eqref{bc_2} from the TW frame to the slab frame. Such a term is absent when one calculates the SHG of the waveguide geometry in isolation (as e.g., in~\cite{Khurgin_SHG_MDM,Kivshar_SHG_MDM}\cite[Ch.~4]{knr_thesis}), meaning that it is \emph{not} intrinsic to the waveguide, but rather, is a result of transforming the TW structure. The $1/(d^2 + \ty^2)$ dependence determines the spatial variation of the SH amplitudes $\mathcal{A}_{sl},~\mathcal{B}_{sl}$ (see Eqs.~\eqref{ansatz_sl}-\eqref{sl_bc2_impl}) along $\ty$ together with the SH source $\Delta_r$; it suppresses SH source strength and thereby, the strength of the SH fields for large $\ty$.
\bigskip

\subsection{Transformation back to the TW frame}\label{sec:back_transfm} 
Our next step is to transform the SH solution $H_{z,sl}^{2\omega}$ back to the TW frame. Since $H_{z,sl}^{2\omega}$ is preserved under the transformation, see Appendix~\ref{apen:Hz_transfrm}, $H_{z,TW}^{2\omega}$ can be obtained by just rewriting the slab coordinates in terms of the TW coordinates. Thus, $H_{z,TW}^{2\omega}$ is given by
\begin{widetext}
\begin{eqnarray}
H_{z,TW}^{2\omega} = 
\begin{cases} 
\mathcal{A}_{TW}(x,y)\exp\left( \frac{4 i a \alpha^\omega |y|}{x^2 + y^2}\right)\sinh\left(\frac{4 a \alpha^\omega x}{x^2 + y^2}\right), \quad  
\textrm{when}~~ x^2 + y^2 + 2a|x| \geq 0\,, \\
\mathcal{B}_{TW}(x,y)\exp\left(\frac{4 i a \alpha^\omega |y|}{x^2 + y^2}\right)\textrm{sgn}[x] \exp\left(\frac{- 4 a \alpha^\omega |x|}{x^2 + y^2}\right) , \textrm{when}~~ x^2 + y^2 + 2a|x| < 0,
\end{cases} \label{sh_sol1} 
\end{eqnarray}
\end{widetext}
where
\begin{gather}
\mathcal{A}_{TW}(x,y) = \frac{-i\omega\varepsilon_0 \varepsilon_{bg}^{2\omega}}{2\alpha^\omega \mathcal{P}} ~\mathcal{C}(x,y), \\
\mathcal{B}_{TW}(x,y) = \mathcal{A}_{TW}(x,y)~e^{2\alpha^\omega} \sinh 2\alpha^\omega , \label{sh_sol_param1} \\
\mathcal{C}(x,y) = \frac{4a(x^2 + y^2)^2}{4a^2y^2 + (x^2+y^2)^2}~ \mathcal{R} \left(\tau_x,\tau_y\right)\,, \label{sh_sol_param2} \\
\tau_x=\frac{ 1/(2a)}{1/(4a^2) + y^2/(x^2+y^2)^2 }, \label{tau_x}\\
\tau_y=\frac{ y/(x^2+y^2)}{1/(4a^2) + y^2/(x^2 + y^2)^2}. \label{tau_y}
\end{gather}

\begin{figure*}
\centering
\includegraphics*[width=0.7\textwidth]{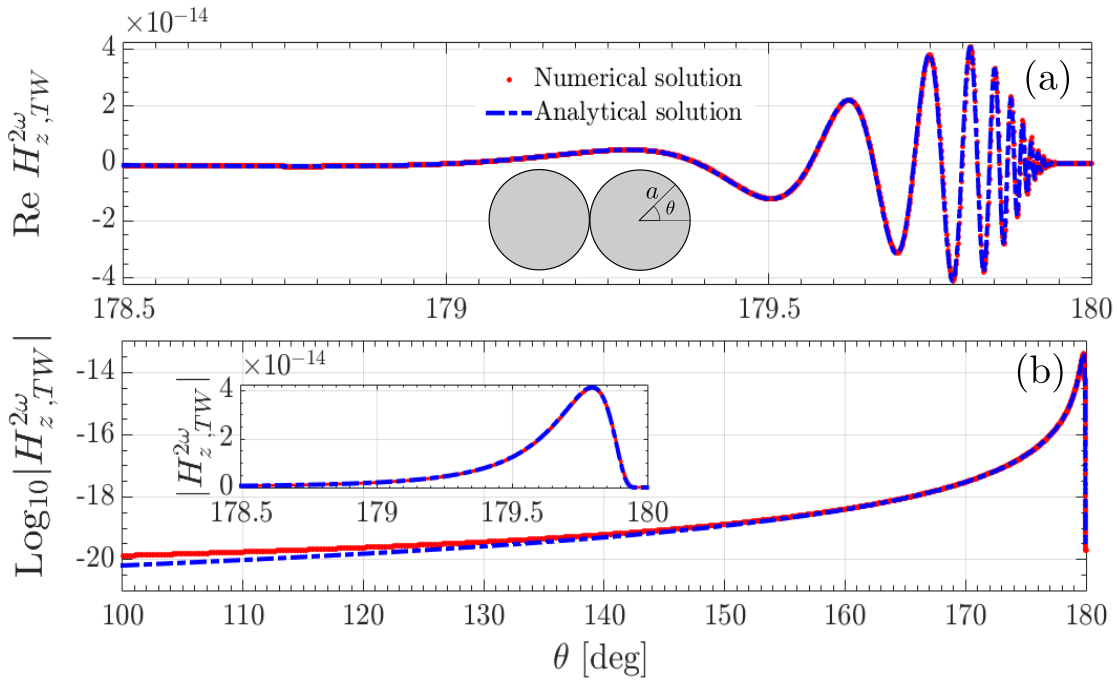} 
\caption{(Color online) (a) SH solution $H_{z,TW}^{2\omega}$ on the circumference of the right wire (close to the touching point) as the function of angle $\theta$. The dashed-dot curve is the analytical solution as evaluated by Eq.~\eqref{sh_sol_simp} and the dotted curve is the numerical solution. The SH wavelength is chosen to be $500$nm (with FF wavelength of $1000$nm) and the other parameters used are $a = 5$ nm, $\varepsilon_{bg}^\omega = \varepsilon_{bg}^{2\omega} = 1$, $\varepsilon_m^\omega = -48.6 + i8.65$, $\varepsilon_m^{2\omega} = -8.67 + i1.10$, $\chi^{(2)}_{S,\perp\perp\perp} = 10^{-20}$ m$^2$/V and $E^{\omega}_{0x} = 1$V/m. (b) Same as in Figure~\ref{fig_sh_sol}(a) for a wider angular range. }
\label{fig_sh_sol} 
\end{figure*}
Comparing the spatial dependence of $\mathcal{R}$ in Eq.~\eqref{jmz_decmpr} to Eq.~\eqref{sh_sol_param2} reveals that $\mathcal{R}$ now gains a different spatial dependence denoted by $\tau_{x,y}$; $\tau_{x,y}$ in Eqs.~\eqref{tau_x}-\eqref{tau_y} map every point in the TW plane ($x-y$ plane) onto the circumference of the right wire while, the circumference of the right wire (i.e., when $x^2 + y^2 = 2 a x$) is mapped onto itself, i.e., $\tau_x=x,~\tau_y = y$. This map is a consequence of Eq.~\eqref{ansatz_sl} and Eq.~\eqref{sl_bc2_impl} as the amplitude $\mathcal{A}_{sl}$ in the slab frame is determined by the spatial variation of $\Delta_r$ at $\tx=d$. 

We now focus on the spatial variation of $H_{z,TW}^{2\omega}$ on the circumference of the right wire. Since this curve maps onto itself, the functional dependence of $\mathcal{R}(\tau_x,\tau_y)$ in Eq.~\eqref{sh_sol_param2} simplifies to $\mathcal{R}(x,y)$. Also the additional pre-factor in Eq.~\eqref{sh_sol_param2} simplifies to $2x$. Thus, $H_{z,TW}^{2\omega}$ is given by
\begin{eqnarray*}
H_{z,TW}^{2\omega} (x,y) =&-i&\omega\varepsilon_0\varepsilon_{bg}^{2\omega} \left( \frac{ \sinh 2\alpha^\omega}{\alpha^\omega \mathcal{P}  }\right) x~\mathcal{R}(x,y) \\ &\times& \exp\left( \frac{ 4i a \alpha^\omega |y|}{x^2 + y^2}\right),\nonumber
\end{eqnarray*}
and upon using Eq.~\eqref{jmz_decmpr}, the above can be rewritten as
\begin{equation}
H_{z,TW}^{2\omega}(x,y) = - i \omega \varepsilon_0 \varepsilon_{bg}^{2\omega} \left(\frac{ \sinh 2\alpha^\omega}{\alpha^\omega \mathcal{P}} \right)~x~ J_{z,r}(x,y). \label{sh_sol_simp}
\end{equation}
This analytic solution is the main result of the paper.

In order to validate our analytical solution, we compare it with a numerical solution, obtained using a commercially-available finite element method software package COMSOL Multiphysics (see Appendix~\ref{apen:COMSOL} for further details). Figure~\ref{fig_sh_sol}(a) shows an excellent agreement between the analytic and numeric solutions of $H_{z,TW}^{2\omega}$ on the circumference of the right wire close to the touching point ($\theta = 180^{\circ}$). In Figure~\ref{fig_sh_sol}(b), we see that this agreement extends even to angles as large as several tens of degrees away from the touching point. This validates our analytical solution.  

Similar to the linear response, the SH response exhibits interesting physical phenomena like wavelength compression, slow light, field enhancement and energy accumulation close to the touching point~\cite{alex_kissing_NJP}, the latter being inhibited by the absorption. In what follows, we interpret the SH solution of the TW, and highlight how the terms that appeared in the solution for the slab manifest themselves in the solution for the TW geometry and how they affect the above mentioned phenomena.

First, the momentum mismatch between the mode and the SH source is preserved under the transformation, showing that (unlike previous claims in the literature (e.g.,~\cite{zayat_rev})), the phase-matching condition is relevant even for this sub-wavelength structure~\footnote{For the set of parameters considered in Figure~\ref{fig_sh_sol}, $|\mathcal{P}|\approx 1$. This implies that the SHG indeed occurs under phase-mismatched condition.}. Indeed, the link between the TW geometry and the slab geometry revealed by the coordinate transformation shows that if PM is important for the slab geometry, then, it must be important also for the particle (i.e., TW) geometry. Such an effect is possible due to the (non-uniform) wavelength compression induced by the (touching) wires which makes the optical length along the wires effectively infinite~\cite{Alex_kissing_cyls_NL,alex_kissing_NJP}. We note that such a result is nontrivial to reproduce by a multipolar expansion, despite the deep subwavelength nature of the TW geometry~\cite{Selection_rules}. Second, MM manifests through the symmetry of the SH sources for the identical TW system. The $J_{z,r/l}$ is anti-symmetric in nature (see Figure~\ref{fig_jmz}(c)), thus generating an anti-symmetric $H_{z,TW}^{2\omega}$ field.

Apart from the phase-matching and mode-matching conditions, the SH solution also exhibits an additional spatial dependence which occurs as the prefactor in $\mathcal{C}$ (see Eq.~\eqref{sh_sol_param2}). We coin this term as the \emph{geometric factor}. The geometric factor stems from transforming  $1/(d^2 + \ty^2)$ in Eq.~\eqref{sl_bc2_impl} to the TW frame. Recall that $1/(d^2 + \ty^2)$ was obtained from transforming generalized boundary condition from TW to slab frame, however, its signature still remains in the final solution as we transform only the SH magnetic field back from the slab to the TW frame leaving out the transformation of generalized boundary condition~\eqref{hnztr}.

As mentioned, the geometric factor on the circumference of the TW simplifies to $2x$, i.e., it attains small values close to the touching point ($x \approx 0$). Since the magnetic field solution (see Eq.~\eqref{sh_sol_simp}) is given by the product of the SH source and the geometric factor, the SH source (and thereby, the SH response) is further suppressed by the geometric factor close to the touching point. Therefore, the SH solution decays faster to the touching point compared with the linear solution.

The origin of the geometric factor can be understood as follows. In the usual Green's function approach, a distributed source should be integrated over its spatial coordinates along with the Green's tensor corresponding to the structure, namely, $\int\bar{\bar{G}}_{TW}(\mathbf{r},\mathbf{r}') \mathbf{P}^{2\omega}(\mathbf{r}') d \mathbf{r}'$. This integral contains all the information about the structure (in particular, the modal structure) and the spatial distribution of the SH source, and the convolution naturally yields PM and MM. However, prior to the evaluation of this integral, the final spatial distribution of the SH solution is not known. The conformal transformation reveals that the SH solution~\eqref{sh_sol1}-\eqref{tau_y} has a remarkably simple form, namely, it is linearly proportional to the SH source, phase-matching and mode-matching factors. Additionally, we encounter the geometric factor which can be thought of as a simplification of the Green's function integral with the distributed source corresponding to the TW geometry. We note that such a simplification/factorization of SH solution is realizable due the power of conformal transformation by transforming the seemingly difficult SH problem to the much simpler system, i.e., the slab frame and thereby, enabling us to relate the local SH solution in terms of the the local source. Yet, this simplicity pertains only exactly at the metal-dielectric interface of the TW, and is far more complicated elsewhere.

\section{Discussion and Outlook}\label{sec:diss_out}

Using the technique of conformal transformations, we have related the SHG in waveguides to the SHG in a far more complicated geometry - the touching wire dimer. Our ansatz approach is simpler compared to the treatment of the linear TW problem with CT, as we avoid the need to transform to momentum space and perform contour integrations. The transformation allows us to unfold the hidden symmetries, separating variables such that the roles of PM and MM become obvious also in the TW frame, making the interpretation of the final solution quite straightforward. 

In particular, this showed the equivalence of phase-matching in the slab geometry to tuning to the localized surface plasmon resonance at the SH for the TW structure. However, our analysis shows rather surprisingly, that PM and MM and source strength are accompanied by an additional factor. This additional factor was not identified before and we refer to it as a geometric factor. Indeed, its presence implies that different original geometries might give rise to different sources, hence different geometric factors. This shows that the solution for the particle geometry exhibits richer physics compared to the slab. In particular, the revealed complexity demonstrates the limitations of approaches based on just evaluating the strength of the source $P^{2\omega}$, or just on PM~\footnote{In~\cite{pavel_zayats_PRB_mode_matching}, the SHG efficiency was optimized via an integral over the {\em longitudinal} coordinate involving the surface charges. While in~\cite{pavel_zayats_PRB_mode_matching} it was referred to as mode matching, adopting the link between the particle and waveguide geometries revealed in this manuscript, we argue that it in fact represents phase matching instead.}.

This work introduces new tools and insights into nonlinear optical wave mixing for nanoscale structures (and on the treatment of distributed sources on the nanoscale \cite{thermalreview,casimirreview,casimircalculations}). It differs from many studies by providing a unique analytic near-field solution rather than on a qualitative description of the (experimentally-accessible) far-field pattern% e.g., Roke-Bonn-Petrushev (also limited to PWs)
, and by going beyond qualitative symmetry breaking arguments~\cite{NL_scattering_Roke_Bonn_Petrushev,symmetry_break_Zyss,symm_brk_Martti,symm_breaking_Miano}.

The solution approach adopted in this manuscript can serve as a means to study and optimize the near-field and the spectral response of the touching wire dimer. It can also be used to calculate the SH field distributions analytically for various other singular nanoparticle structures like asymmetric touching wires, crescent structures, wedge structures, circular protrusion from a planar interface, etc.~\cite{TO_vdw_science}. Additionally, this approach can also be extended to calculate the SH response from non-singular structures such as non-touching wires~\cite{Kurt_SHG_Bulk_TO}, blunt crescent structures~\cite{Yu_bluntness}.

The solution approach adopted in this manuscript is suitable also for other elements of the surface polarization tensor as well as for (non-local) bulk polarizations (by mapping them to a surface polarization~\cite{sipe_srf_map,miano,ciraci_prb_SHG,ciraci2}).

Other problems that can be solved with the same approach are 3D structures~\cite{Antonio_kissing_spheres_PRL,TO_vdw_PNAS}, SHG at other spectral regimes, e.g., in the THz regime~\cite{THz_SHG}, optical rectification, and more complex nonlinear wave interactions, such as 3 wave mixing~\cite{Tal_nat_phot_SH_metasurface,Tal_SH_beam_shaping_ACS,Ponomarenko_SFG}, phase-conjugation~\cite{tr_super_lens_pendry} or even additional effects such as the role of non-locality~\cite{Antonio_kissing_cyls_nonlocal_PRL,Antonio_kissing_cyls_nonlocal_PRB} on SH and many more.

\begin{acknowledgements}
The authors would like to thank A. Niv and A. Isha'aya for many useful discussions. KNR and YS were partially supported by Israel Science Foundation (ISF) grant (899/16). KNR and AIFD would like to acknowledge STSM Grant from the COST Action MP1403. YS acknowledges the financial support from the People Programme (Marie Curie Actions) of the European Union’s Seventh Framework Programme (FP7/2007-2013) under REA grant (333790) and the Israeli National Nanotechnology Initiative. AIFD acknowledges funding from EU Seventh Framework Programme under Grant Agreement FP7-PEOPLE-2013-CIG-630996, and the Spanish MINECO under contract FIS2015-64951-R.
\end{acknowledgements}

\appendix
\section{Extracting the magnetic fields in the quasi-static regime} \label{apen:lin_mag}

The linear electric fields were obtained by setting $\nabla\times\mathbf{E}^\omega=0$~\cite{Alex_kissing_cyls_NL}, thus, the linear magnetic field was set to zero. For the quasi-static structures, the electric response dominates. However, the magnetic response even though negligible, is not truly zero. Here, we demonstrate on how to extract the magnetic field from the quasi-static structures.

In general, this magnetic field can be evaluated in two different ways. One can express the excitation sources in terms of the magnetic sources and solve for the magnetic response. For example, a plane wave illumination was modelled using magnetic line currents and the magnetic response for various quasi-static structures was obtained~\cite{Yang+Paloma_2018}. Similarly, in this manuscript, we formulate the nonlinear polarization in terms of a surface magnetic current to obtain the SH magnetic response of the TW system. Alternatively, despite setting $\nabla\times\mathbf{E}=0$, one can employ Ampere's law to extract the magnetic field from the evaluated electric fields. In what follows, we extract the linear magnetic field $H_{z,TW}^\omega$ of the TW system under plane wave illumination using the latter. 

Using Ampere's law explicitly for our case yields 
\begin{subequations}
\begin{align} \label{ampr1} 
\partial_{x}H_{z,TW}^\omega(x,y) &= i \omega \varepsilon_0 \varepsilon_{bg/m}^\omega E_{y,TW}^{\omega}(x,y), \\
\partial_{y}H_{z,TW}^\omega(x,y) &= - i \omega \varepsilon_0 \varepsilon_{bg/m}^\omega E_{x,TW}^{\omega}(x,y), \label{ampr2} 
\end{align}
\end{subequations}
with $\varepsilon_{bg/m}^\omega$ considered in the appropriate domains. $E_{x,TW}^{\omega}$ and $E_{y,TW}^{\omega}$
in Eqs.~\eqref{ampr1}-\eqref{ampr2} correspond to the linear electric field solutions of the identical TW system in different domains as obtained by Lei \textit{et al.}~\cite{alex_kissing_NJP}\footnote{Note that Lei \textit{et al.} obtained the linear response for $\varepsilon_{bg}^\omega = 1$. However, one can extend this linear response to arbitrary $\varepsilon_{bg}^\omega$ by substituting $\varepsilon_m^\omega\rightarrow \varepsilon_m^\omega/\varepsilon_{bg}^\omega$}.
Upon integrating Eqs.~\eqref{ampr1}-\eqref{ampr2} and determining the constants of integration we arrive at
\begin{widetext}
\begin{eqnarray}
H_{z,TW}^\omega(x,y)=
\begin{cases}
\mathcal{N}~\textrm{sgn} [y] ~\cosh\left( \frac{2a\alpha^{\omega} x}{x^2+y^2}\right)\exp{\left(\frac{2ia\alpha^{\omega} |y|}{x'^2+y'^2}\right)}, \quad \textrm{when} \quad x^2+y^2+2|x|a > 0, \\
\left(\frac{\mathcal{N}\varepsilon_m^\omega}{\varepsilon_m^\omega + \varepsilon_{bg}^\omega}\right)\textrm{sgn} [y]\exp\left(\frac{-2a\alpha^{\omega} |x|}{x^2 + y^2}\right) \exp{\left(\frac{2ia\alpha^{\omega} |y|}{x^2 + y^2}\right)}, \quad \textrm{when}\quad x^2 + y^2 +2|x| a < 0,
\end{cases}
\label{lin_Hz_TW}
\end{eqnarray}
\end{widetext}
with $\mathcal{N}=-2\pi i\omega\varepsilon_0 a E^{\omega}_{0x}$. 
\begin{figure}[b!]
\begin{center}
\includegraphics[width=\columnwidth]{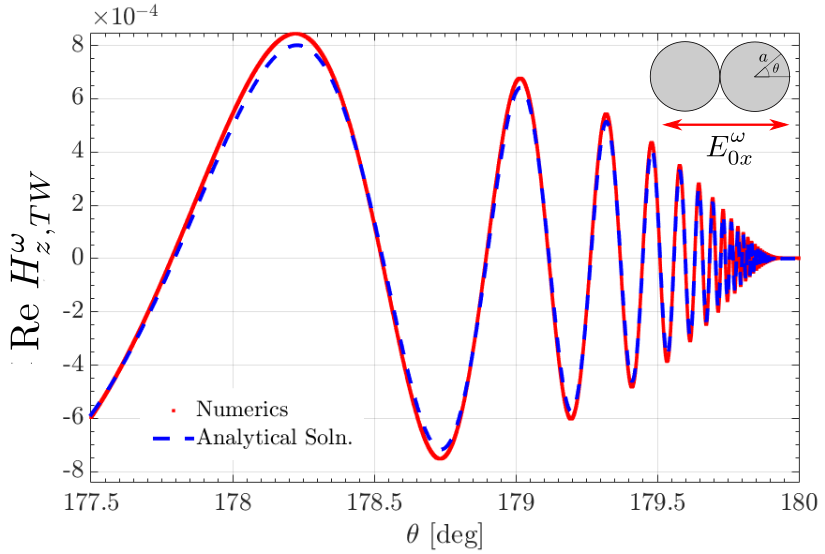}
\caption{Linear magnetic field $H_{z,TW}^\omega$ as the function of angle $\theta$ on the circumference of the right wire. The analytical solution (as evaluated by Eq.~\eqref{lin_Hz_TW}) and the numerical solution are shown in blue and red curves, respectively, at FF wavelength of $500$ nm.The other parameters are $a=5$ nm, $\varepsilon_m^\omega = - 8.3 + 0.29i$, $\varepsilon^\omega_{bg} = 1$ and $E_{0x}^{\omega} = 1$ V/m.}
\label{fig_lin_Hz} 
\end{center}
\end{figure}

Alternatively, one can also arrive at magnetic field $H_{z,TW}^\omega$  by obtaining the magnetic field in the slab geometry can then transforming to the TW frame. The out-of-plane component of the magnetic field is preserved under the 2D conformal transformation (see Appendix \ref{apen:Hz_transfrm}) \cite{Yang+Paloma_2018}. We have verified that both the methods are consistent.

Figure~\ref{fig_lin_Hz} shows the comparison between the analytical and numerical solutions of $H_{z,TW}^\omega$ on the circumference of the right wire of TW system. The numerical solution was obtained using COMSOL Multiphysics by solving for electrodynamic response from TW under plane wave illumination. We find that the analytical solution is in excellent agreement with the numerical solution.

\section{Transformation of the magnetic field} \label{apen:Hz_transfrm}
Here, we derive the transformation rules of the magnetic field under 2D conformal transformation. It is important to note that the transformation considered here is completely in-plane, i.e., 2D, thus, the out-of-plane coordinate $z$ remains unaffected. Following the TO formulas for a vector quantity~\cite{Ward_Pendry_1996}, the magnetic fields in the TW frame $\mathbf{H}_{TW}$ and the slab frame $\mathbf{H}_{sl}$ are related as
\begin{equation}
\begin{pmatrix}
H_{x,TW}(x,y) \\ H_{y,TW}(x,y) \\ H_{z,TW}(x,y)
\end{pmatrix} = \begin{pmatrix}
\Lambda_{11} & \Lambda_{12} & 0 \\ 
 \Lambda_{21} & \Lambda_{22} & 0 \\ 
0 & 0 & 1
\end{pmatrix} 
\begin{pmatrix}
H_{\tx,sl}(\tx,\ty) \\ H_{\ty,sl}(\tx,\ty) \\ H_{z,sl}(\tx,\ty)
\end{pmatrix},
\label{a2e1} 
\end{equation}
where $\Lambda_{ij}$ is the Jacobian of the transformation. Equating the $z$-components in Eq.~\eqref{a2e1} we arrive at
\begin{equation}
H_{z,TW}^{2\omega}(x,y) = H_{z,sl}^{2\omega}(\tx(x,y),\ty(x,y)). \label{a2e2} 
\end{equation}
Thus, we have obtained that the $z$-component of the magnetic field is invariant under the transformation~\cite{Yang+Paloma_2018}. This conclusion holds true for both FF and SH magnetic fields.

It also follows from Eqs.~\eqref{a2e1}-\eqref{a2e2} that any vector pointing normal to the $x-y$ plane or $\tx-\ty$ plane, for example, SH source $J_{z,r/l}$, is preserved under 2D conformal transformation.

\section{Amplitudes of SH source - $\mathcal{R}$ and $\mathcal{L}$}\label{apen:RandL}
The amplitudes of the SH sources $\mathcal{R}$ and $\mathcal{L}$ for the TW as defined in Eqs.~\eqref{jmz_decmp} can be evaluated as follows. The FF response $E_{\perp}^{\omega}$ at the metal-dielectric interface on the right and left cylinders(on the metal side) is given by~\cite{alex_kissing_NJP}
\begin{subequations}
\begin{align}
E^{\omega}_{\perp,r}(\theta_r) &= \mathcal{K} \left[ \frac{ie^{i\left|\theta_r\right|}}{\left(1+e^{i\left|\theta_r\right|}\right)^2}\exp\left( \frac{-\alpha^{\omega}}{1+e^{i\left|\theta_r\right|}} \right)\right],\\
E^{\omega}_{\perp,l} (\theta_l)&= \mathcal{K} \left[ \frac{i\mathcal{L}(\theta_l)}{\left(1-\mathcal{L}(\theta_l)\right)^2}\exp\left( \frac{-\alpha^{\omega}}{1-\mathcal{L}(\theta_l)} \right)\right],    
\end{align}
\label{RandL}
\end{subequations}
respectively. Other variables in Eqs.~\eqref{RandL} are defined as
\begin{widetext}
\begin{eqnarray}
\mathcal{K}&=&\left(\frac{\pi\alpha^{\omega}\varepsilon^{\omega}_{bg}}{ \varepsilon^{\omega}_{m}+\varepsilon^{\omega}_{bg}}\right)E_{0x}^{\omega},  \nonumber \\
\theta_r &=& \tan^{-1}\left( \frac{y}{x-a} \right), \textrm{where} ~ \left\lbrace \left.(x,y)\in \mathbf{R} ~ \right| x^2+y^2-2ax=0\right\rbrace, \label{thr} \\
\theta_l &=& \tan^{-1}\left( \frac{y}{x+a} \right), \textrm{where} ~ \left\lbrace \left.(x,y)\in \mathbf{R} ~ \right| x^2+y^2+2ax=0\right\rbrace, \label{thl}\\
\mathcal{L}(\theta_l)&=&\begin{cases} e^{-i\theta_l}~~ \forall ~\theta_l \in [0,\pi), \\
e^{i\theta_l} ~~\forall ~\theta_l \in [\pi,2\pi). \nonumber
\end{cases}
\end{eqnarray}
The variable $\theta_r$ ($\theta_l$) in Eq.~\eqref{thr} (Eq.~\eqref{thl}) corresponds to the angular co-ordinate on the circumference of the right (left) cylinder of radius $a$ centered at $x=a$ ($x=-a$). The SH source~\eqref{jmz} on the right (left) cylinder in terms of $E_{\perp,r}^{\omega}$ ($E_{\perp,l}^{\omega}$) is given by 
\begin{subequations}
\begin{align}
J_{z,r}(\theta_r)&=\frac{\chi^{(2)}_{\perp\perp\perp}}{a\varepsilon^{\omega}_{bg}}~ \partial_{\theta_r}\left[E^{\omega}_{\perp,r}(\theta_r)\right]^2, \label{jr_explct}\\  
J_{z,l}(\theta_l)&=\frac{\chi^{(2)}_{\perp\perp\perp}}{a\varepsilon^{\omega}_{bg}}~\partial_{\theta_l}\left[E^{\omega}_{\perp,l}(\theta_l)\right]^2, \label{jl_explct}
\end{align}
\label{j_explct}
\end{subequations}
respectively. In writing Eqs.~\eqref{j_explct} we have used the fact that the tangential derivative $\partial_{\parallel}$ on the right (left) cylinder is given by $\frac{1}{a}\partial_{\theta_r}$ ($\frac{1}{a}\partial_{\theta_l}$). Rewriting $J_{z,r}$ and $J_{z,l}$ (in Eqs.~\eqref{j_explct}) in ($x,y$) and invoking the decomposition of $J_{z,r/l}$ from Eq.~\eqref{jmz_decmp} gives us $\mathcal{R}$ and $\mathcal{L}$ as
\begin{subequations}
\begin{align*}
\mathcal{R}(x,y)&=J_{z,r}(x,y)\exp\left(\frac{-4ia\alpha^{\omega}|y|}{x^2+y^2}\right), \textrm{where} ~ \left\lbrace \left.(x,y)\in \mathbf{R} ~ \right| x^2+y^2-2ax=0\right\rbrace, \\  
\mathcal{L}(x,y)&=J_{z,l}(x,y)\exp\left(\frac{-4ia\alpha^{\omega}|y|}{x^2+y^2}\right), \textrm{where} ~ \left\lbrace \left.(x,y)\in \mathbf{R} ~ \right| x^2+y^2+2ax=0\right\rbrace,
\end{align*}
\end{subequations}
respectively.
\end{widetext}
\section{Transformation of the generalized boundary condition to the slab frame} \label{apen:gnbc_transfrm}
Here, we compute the transformation of the generalized boundary condition~\eqref{full_gnbc} from TW to slab frame. In what follows, we first compute the transformation of the left-hand-side of Eq.~\eqref{full_gnbc} and then followed by its right-hand-side. The magnetic field $H_{z,TW}^{2\omega}$ can be replaced by $H_{z,sl}^{2\omega}$ (see Appendix~\ref{apen:Hz_transfrm}). Replacing the spatial coordinates and the derivatives of TW frame by that of the slab ones shows that the left-hand-side of Eq.~\eqref{full_gnbc} transformations as
\begin{equation}
\left(F_1\partial_{\tx }+ F_2\partial_{\ty}\right) \left[ \frac{H_{z,sl}^{2\omega}|_{bg}}{\varepsilon_{bg}^{2\omega}} - \frac{H_{z,sl}^{2\omega}|_{m}}{ \varepsilon_m^{2\omega}}\right] , \label{derv_transf} 
\end{equation}
where
\begin{equation*}
F_1 = \frac{\pm(\tx^2-\ty^2)-2d\tx}{g^2}, \quad F_2 = \pm \frac{2\ty(\tx\mp d)}{g^2}. 
\end{equation*}
We now compute the transformation of the right-hand-side of Eq.~\eqref{full_gnbc}. The magnetic current $J_{z,r/l}$ is a vector normal to the $x-y$ plane, thus, it is preserved under the transformation. Invoking the decomposition of $J_{z,r/l}$ from Eqs.~\eqref{jmz_decmp} and rewriting it in terms of the slab variables gives
\begin{subequations}
\begin{align}
J_{z,r}&\rightarrow\Delta_r(\tx,\ty)~e^{2i\alpha^{\omega}|\ty|/d} ~\delta \left(\tx- d\right), \label{delr} \\
J_{z,l}&\rightarrow\Delta_l(\tx,\ty)~e^{2i\alpha^{\omega}|\ty|/d} ~\delta \left( \tx+ d \right). \label{dell} 
\end{align}
\label{del}
\end{subequations}
In writing the above expressions, we have relabeled $\mathcal{R}$ and $\mathcal{L}$ in slab geometry by
\begin{subequations}
\begin{align}
\Delta_r(\tx,\ty)&\equiv \mathcal{R} \left( \frac{g^2\tx}{\tx^2+\ty^2},~ \frac{g^2\ty}{\tx^2+\ty^2}\right), \\
\Delta_l(\tx,\ty)&\equiv\mathcal{L} \left( \frac{g^2\tx}{\tx^2+\ty^2},~ \frac{g^2\ty}{\tx^2+\ty^2}\right),
\end{align}
\end{subequations}
respectively. Since the conformal transformation~\eqref{eq_CT} preserves the symmetry relation between $\mathcal{R}$ and $\mathcal{L}$, we have $\Delta_r(\tx=d,\ty) = - \Delta_l(\tx = - d,\ty)$. It can seen from Eq.~\eqref{delr} (Eq.~\eqref{dell}) that the surface source $J_{z,r}$ ($J_{z,l}$) on the circumference of the right (left) wire transforms as the source placed at right (left) interface of the slab geometry. 
 
Thus, by combining Eqs.~\eqref{derv_transf}-\eqref{del}, we arrive at the transformed generalized boundary condition given by
\begin{eqnarray}
\left(\frac{\pm(\tx^2-\ty^2)-2d\tx}{g^2}\right){\partial}_{\tx}\left[ \frac{H_{z,sl}^{2\omega}|_{bg}}{\varepsilon_{bg}^{2\omega}} - \frac{H_{z,sl}^{2\omega}|_{m}}{ \varepsilon_m^{2\omega}}\right]  \nonumber \\
=- 2 i \omega \varepsilon_0~\Delta_{r/l}(\tx=\pm d,\ty)~e^{2i\alpha^{\omega}|\ty|/d} \delta	(\tx\mp d).
\end{eqnarray}
%%%%%%%%%%%%%

\section{Numerical Simulations} \label{apen:COMSOL}
The numerical solutions are obtained using the commercially available finite element method software package COMSOL Multiphysics 3.5a. We have used harmonic propagation analysis in the radio frequency module to obtain the SH electrodynamic response of the TW system. The FF analytical expressions were used to compute the SH source $J_{z,r/l}$~\eqref{jmz} and were given as an input to the COMSOL 3.5a solver. Specifically, SH source $J_{z,r/l}$ was inserted as a source in the `Boundary setting - Equation system' node. The region close to the touching point was resolved with mesh sides below $10^{-4}$ nm to sample the rapidly oscillating $J_{z,r/l}$ accurately and to tackle the geometric singularity. A perfectly matched layer was used to suppress the reflections from the boundaries of the simulation domain. The convergence of the numerical solution as a function of mesh size and size of the simulation domain was verified. Despite the availability of the latest versions (COMSOL 5.3a), we were forced to resort to the old version COMSOL 3.5a due to its lower sensitivity to the ill-conditioning associated with the extremely fine mesh near the touching point. A detailed account concerning the sensitivity of COMSOL 5.3a to the small mesh and evaluation of the linear and SH response is presented in Ref.~\cite[Appendix~D]{knr_thesis}.

%\bibliography{ref}
%%%%%%%%% Copied from .bbl file
%%%%%%%%% Copied from .bbl file
%apsrev4-2.bst 2018-12-27 (MD) hand-edited version of apsrev4-1.bst
\providecommand{\noopsort}[1]{}\providecommand{\singleletter}[1]{#1}%

\end{document}